# Three-dimensional analysis of annual layers in tree trunk and tooth cementum

Bert Müller*[a,b], Muriel Stiefel[a], Griffin Rodgers[a], Mattia Humbel[a], Melissa Osterwalder[a], Jeannette von Jackowski[a,b], Gerhard Hotz[c,d], Adriana A. Velasco[e], Henry T. Bunn[e], Mario Scheel[f], Timm Weitkamp[f], Georg Schulz[a,g], and Christine Tanner[a]

[a]Biomaterials Science Center, Department of Biomedical Engineering, University of Basel, Gewerbestrasse 14, 4123 Allschwil, Switzerland; [b]Biomaterials Science Center, Department of Clinical Research, University Hospital Basel, Schanzenstrasse 55, 4051 Basel, Switzerland; [c]Natural History Museum of Basel, Anthropological Collection, 4501 Basel, Switzerland; [d]Integrative Prehistory and Archaeological Science, University of Basel, 4055 Basel, Switzerland; [e]Department of Anthropology, University of Wisconsin-Madison, Madison, Wisconsin, USA [f]Synchrotron Soleil, 91192 Gif-sur-Yvette, France; [g]Core Facility Micro- and Nanotomography, Department of Biomedical Engineering, University of Basel, 4123 Allschwil, Switzerland

**ABSTRACT**

The growth of plants, animals, and humans can give rise to layered structures associated with annual periodicity. Thickness variations are often correlated to nutrition supply and stress factors. The annual layers in a tree trunk with millimeter thickness can be directly counted, whereas the layers in tooth cementum with micrometer thickness are made visible using optical microscopy. These optical techniques rely on the surface evaluation or thin, optically transparent slices. Hard X-ray tomography with micrometer resolution, however, provides a three-dimensional view without physical slicing. We have developed a procedure to enhance the tomography data of annual layers in human and bovid tooth cementum. The analysis of a substantial part of an archeological human tooth demonstrated that the detected number of layers depended on the selected region and could vary between 13 and 27. The related average thickness of the annual layers was found to be (5.4 ± 1.9) µm for the human tooth, whereas the buffalo tooth exhibited a layer periodicity of 46 µm. The present study elucidates the potential of combining computational tools with high-quality micro computed tomography data to quantify the annual layers in tooth cementum for a variety of purposes including age-at-death determination.

## 1. INTRODUCTION

Layered structures are frequently found in nature. One of the best-known examples are tree's annual growth rings. Their width is a top-notch biological indicator on dry seasons, excessive rain, air pollution and light conditions, just to mention a few parameters. Therefore, the estimation of the tree's age from its diameter is error prone, especially if the local growth conditions are unknown. The physical cutting of the trunk makes the information easily accessible by visual inspection. Besides the average layer thickness, the intra- and interlayer thicknesses provide a wealth of quantities on the local climate and light condition changes, see Fig. 1.

Computed tomography (CT) is a three-dimensional imaging technique, which enables us to visualize the annual layers in wood without physical slicing, see for example [1]. It is, therefore, especially supportive for isotropic imaging with micrometer resolution in the three orthogonal directions. In addition, the prevention of physical slicing is often necessary to keep integrity of unique wooden blocks. Nevertheless, there are only a limited number of studies on the annual trunk layers by means of micro computed tomography (µCT), even though the measurements are straightforward [1] see references therein.

*bert.mueller@unibas.ch; phone 41 61 207 5430; www.bmc.unibas.ch

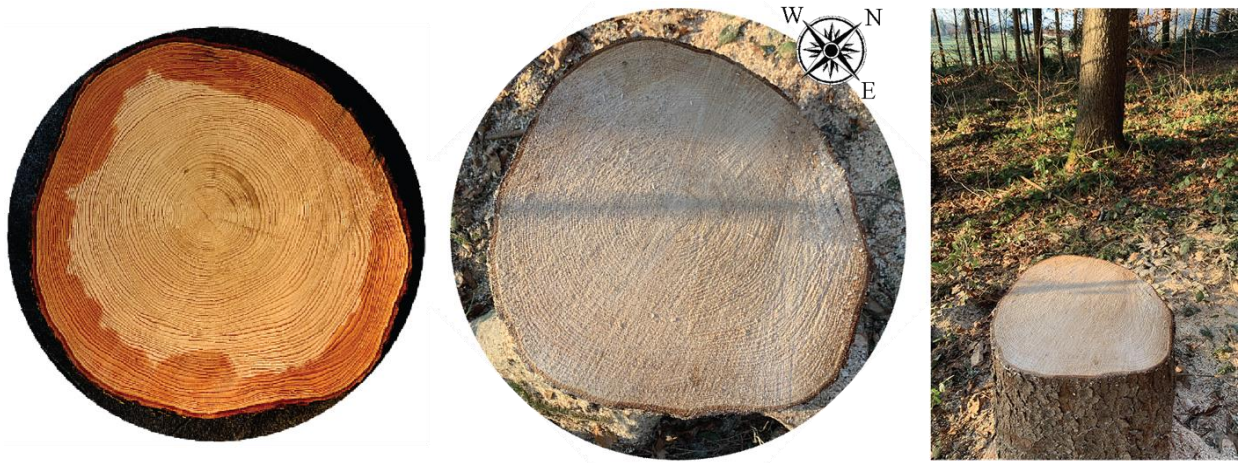

Figure 1. The cross-sectional view on the left clearly demonstrates that this almost 90-year-old tree grew asymmetrically. The thickness variations along the rings indicate changes of the lightening conditions during the growth periods. After cutting down a tree in direct neighborhood, the rings become thicker. The top view on the stump shows the location of the roots, see central photograph. Shadows indicate the geographic direction at position (47.52874° N, 8.59782° O), as also seen on the right photograph taken at January 16, 2022, 3:38 pm with a tree in neighborhood. The maximal diameter of the trunk without bark is 470 mm.

It is much less known that annual layers can be found within the human body. The tooth cementum, *i.e.* the interface between the alveolar bone and the root dentin, contains unique growth features as recognized in optical micrographs of tooth sections in the 1950s [2]. The preparation of these slices is based physical sectioning, a procedure to be avoided for valuable unique objects including archeological teeth. Micro computed tomography enables us to generate virtual slices without cutting and provides three-dimensional data, which can be virtually sliced in any direction of interest [3]. The question arises how far the spatial resolution and contrast do fit for imaging the annual layers in entire human and bovid teeth by means of hard X-ray tomographic imaging.

## 2. MATERIALS AND METHODS

### 2.1 Tree disc

The wood investigated stems from a spruce fir tree, *lat. picea abies*, grown in the canton Zurich nearby Freienstein, Switzerland (47.52874° N, 8.59782° O), see Fig. 1.

### 2.2 Tree-disc analysis using micro computed tomography

Two wooden disks each with a diameter of 180 mm were imaged using the microtomography system nanotom m® (phoenix|x-ray, GE Sensing & Inspection Technologies GmbH, Wunstorf, Germany). The conventional µCT-system is equipped with a nanofocus tube. It was operated at an acceleration voltage of 140 kVp and a beam current of 60 µA. In order to increase the mean photon energy, a 250 µm-thin copper filter was placed between source and wood. The effective pixel length was set to 65 µm. For the tomographic imaging, we have equiangularly acquired 1,800 radiographs along 360° with an exposure time per radiograph of 12 s.

### 2.3 Archeological human teeth

The study includes two teeth, which originate from the well-documented Basel-Spitalfriedhof collection in Switzerland, stored at the Natural History Museum Basel, from the nineteenth century [4]. More than 2,500 patients of the City Hospital were buried between 1845 and 1868 in that cemetery and partly excavated in 1988 and 1989 by the archeological Bodenforschung Basel-Stadt, Switzerland. These skeletons have served as reference series for methodological developments, which included the analysis of stress patterns, see for example [5, 6].

A valuable maxillary first premolar, ID Z_1584, stems from Maria Magdalena Scherb, who lived from May 4, 1829 to August 26, 1865. She visited the City Hospital for nine times. Maria Magdalena Scherb gave birth to two illegitimate children and died as a result of syphilis. Her life was marked by great poverty and deprivation during the early industrialization in Basel.

A 60 to 80 μm-thin tooth slice from an apical part of the middle third of the tooth root towards the crown, ID Z_436, stems from the maxillary first premolar of Maria Eva Kalchschmidt. She lived from November 26, 1802 to August 20, 1851 and gave birth to seven illegitimate children. In 1851 Kalchschmidt felt seriously ill and was admitted to the Bürgerspital. The doctors noted "patient of medium height, very emaciated and miserable-looking, blond-haired, hardly able to speak because of shortness of breath and pain; her great poverty is apparent at first glance".

### 2.4 Archeological human tooth imaging using synchrotron radiation and optical scanner

The tomographic imaging of the ancient human teeth was performed at the microtomography setup of the ANATOMIX beamline at Synchrotron SOLEIL, Gif-sur-Yvette, France [7]. A filtered white beam with a mean photon energy around 33 keV was selected by implementing 20 μm-thin gold and 100 μm copper at an undulator gap of 5.5 mm. The detector was positioned 50 mm downstream of the tooth. An effective pixel size of 0.65 μm was selected. The detector consisted of a 20 μm-thin LuAG scintillator coupled via a 10× magnifying objective to a Hamamatsu Orca Flash 4.0 V2 scientific CMOS camera with 2048 × 2048 pixels each 6.5 μm wide [8]. The exposure time was 0.1 s per projection. Tomographic acquisition consisted of 9,000 radiographs during continuous rotation over an angular range of 360 degrees. The scan time for a single height step with three off-center rings, covering a volume of 7.35 × 7.35 × 1.33 $mm^3$ and consisting of 262 billion voxels (11,307 × 11,307 × 2048 voxels), was 45 minutes.

To capture the whole tooth, we performed extended-field acquisition with three off-center acquisitions. Neighboring radiographs were stitched together by maximizing cross-correlation in the overlapping regions [9-12]. For the complete premolar study, four height steps were analyzed.

Stitched radiographs were flat- and dark-field corrected. To improve signal-to-noise ratio at the expense of spatial resolution, a Gaussian filter with $\sigma = 0.75$ pixels was applied to the projections [13]. Ring artifacts were removed by low-pass filtering the mean of all projections and subtracting it from the corrected projections [14, 15]. Tomographic reconstruction was performed using the gridrec algorithm [16] and the open-source toolbox tomopy (version 1.4.2) [17]. Note, this reconstruction pipeline is not the standard one used at the ANATOMIX beamline.

The optical images of the tooth section were obtained using the Pannoramic MIDI II transmission light microscope. This scanner enables image acquisition with optical magnifications, 20× or 40×. The latter results in a pixel length of 0.12 μm. Three overview scans and 68 consecutive overlapping local scans were acquired to cover the tooth cementum of the selected slice. The acquisition parameters for each local scan were adapted to avoid oversaturation.

### 2.5 Buffalo tooth

The study includes a left upper (maxillary) first molar of an East African buffalo, *lat. syncerus caffer*, from Lake Eyasi, Tanzania sent to death October 26, 1986. The carcass of this buffalo was obtained from a kill site during ethnoarchaeologic work on the hunter and gathering society, the Hadza, and their hunting practices [18]. The death of the individual was observed. Extracted teeth were dimensioned to determine the age-at-death based on the occlusal wear-crown height method [19], which consist of using knowledge on tooth eruption at stages of a particular taxon's life – in this case buffalo, and the degree of wear expected at each stage. The age-at-death was estimated as nine-years for this individual.

### 2.6 Layer visualization in buffalo's tooth cementum

The buffalo tooth was also imaged at the microtomography setup of the ANATOMIX beamline at Synchrotron SOLEIL, France [7]. For this study, we set the undulator gap to 9.5 mm and implemented 20 μm-thin gold and 200 μm copper to obtain a white beam with mean photon energy close to 40 keV. The propagation distance corresponded to 1 m. An effective pixel size of 6.5 μm was selected. The detector consisted of a 600 μm-thick LuAG scintillator coupled via a 1× magnifying objective to a Hamamatsu Orca Flash 4.0 V2 scientific CMOS camera with 2048 × 2048 pixels [8]. The exposure time was set to 0.05 s per projection. Tomographic acquisition consisted of 5,900 radiographs during continuous rotation over an angular range of 360 degrees. The rotation axis was offset to nearly double the detector's field-of-view [20].

Projections were prepared with a Paganin filter [21] using a length of 97 μm [15]. A double flat-field ring correction was applied [15]. The data were reconstructed using the standard processing pipeline at the ANATOMIX beamline at Synchrotron SOLEIL, France.

# 3. IMAGING LAYERED STRUCTURES IN LIFE SCIENCES USING HARD X-RAY TOMOGRAPHY

## 3.1 Tree disc

The wooden disk with almost 0.5 m in diameter, represented in Fig. 1, has been too large to be imaged within the nanotom m®. Therefore, two disks with 0.18 m in diameter have been extracted, see Fig. 2. The photographic images only show the surfaces of the wooden cylinders with clear structural similarities to the related slice of the tomography data, see Fig. 2. The dark brown layers formed during winter time are stronger X-ray absorbing than the bright layers formed in summer. Branches exhibited even higher X-ray absorption, which correlates with a higher density. Both the optical and X-ray data demonstrate that the layer thicknesses can vary from year to year and can depend on the azimuth. The latter is usually not only related to the weather side but mainly given by the shadowing of neighboring trees.

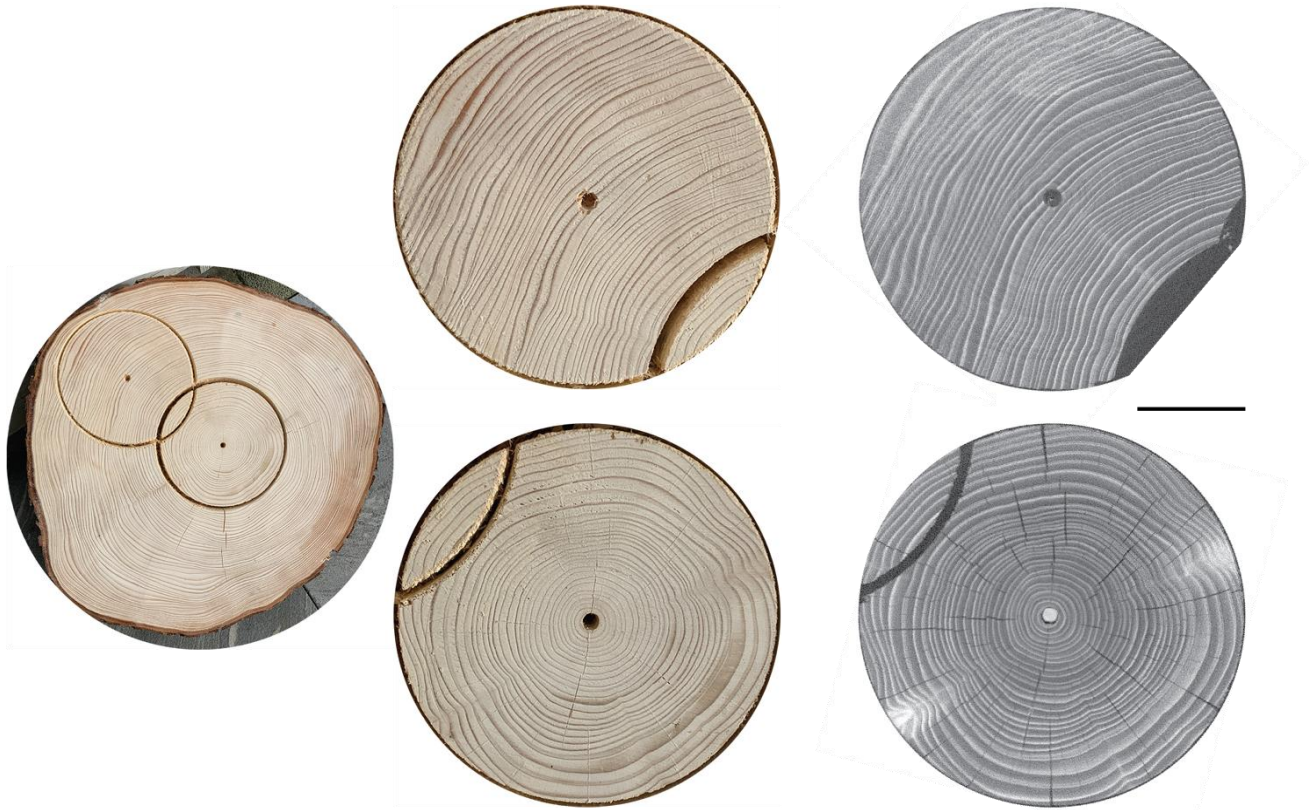

Figure 2. For the tomographic imaging using X rays, two cylinders with a diameter of 180 mm each were prepared, see photographs. The related tomography slices near the surface are represented on the right. The length bar corresponds to 50 mm.

Photographs of the wooden disks are of two-dimensional nature. A three-dimensional representation can only be obtained by physical removal or cutting. The microtomography approach provides a three-dimensional image without physical destruction, see Fig. 3. In this figure, three virtual cuts through the center of mass from one log of wood in the orthogonal directions are displayed. Here, one recognizes four branches. Only two of them are detectable on the surface images given in Fig. 2. They are better visible in the tomography data than on the photographs.

Movies showing a sequence of virtual cuts yield an even better impression on the ring structures, see Video 1. The branches run from the center to the periphery. Their angle to the symmetry axis of the trunk is easily detectable.

Video 2 shows a sequence of tilts of the other wooden disc represented in Fig. 2. At certain angles, part of the annual layers, appear especially intense, because the superposition gives rise to strongest contrast.

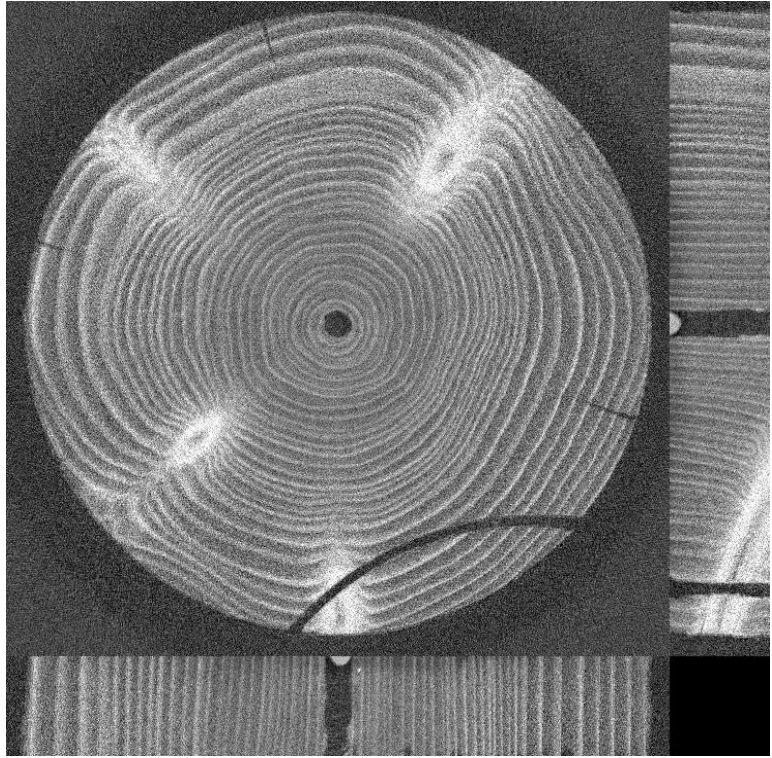

Figure 3. The tomographic data can be virtually cut in any direction. The figure contains the three orthogonal cuts through the center of mass of the cylindrical wooden block.

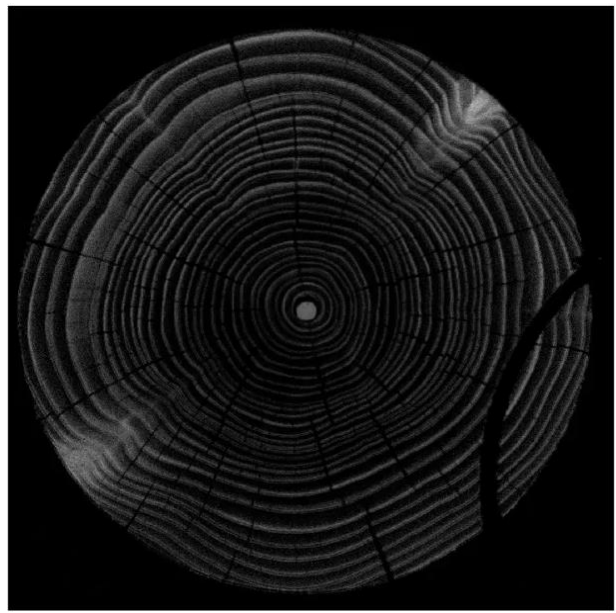

Video 1. Sequence of virtual cuts, which shows the annual rings and the presence of branches:
http://dx.doi.org/doi.number.goes.here

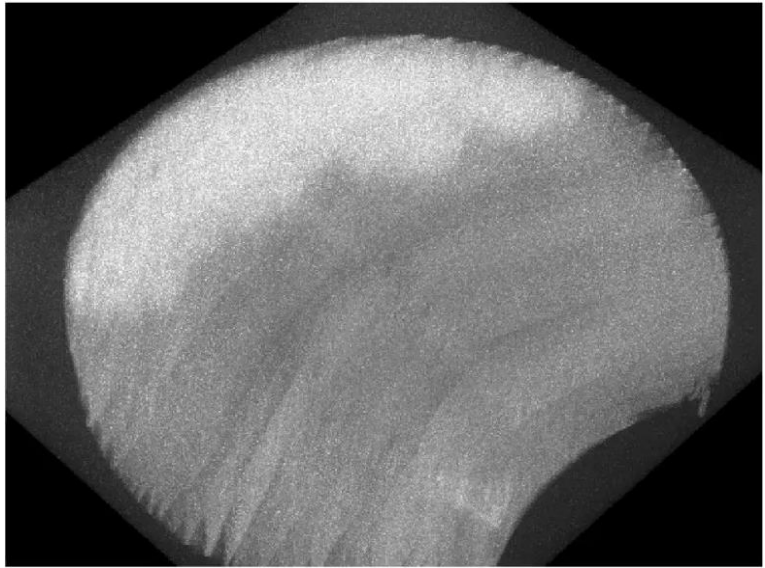

Video 2. Tilting the disk clearly indicates directions for optimized contrast: http://dx.doi.org/doi.number.goes.here

### 3.2 Identification of cementum in an archeological human tooth

Incremental layers are best observed in the acellular cementum, which can be found in the middle third of the tooth root. Therefore, we chose four successive height steps in the middle of the root and analyzed per height step one slab of 35 slices corresponding to 55 µm, see Fig. 4.

The cementum region was manually segmented (by C.T.) to support extraction. The segmentation was performed after integrating the slab in the slice direction. During segmentation, it was found that the cementum boundary on the inner side of the tooth, i.e., at the eruption layer, was generally well visible and hence could be segmented with a high degree of confidence. The outer cementum boundary was less clear. The porous tissue was excluded.

The segmentation results can be seen in blue on the four cross-sections shown in Fig. 4. The cementum is often covered at the periphery of the tooth. Its thickness varies within a height step and from height to height. The mean cementum thickness of the four sections was 91, 85, 95, and 75 µm, with increasing distance from enamel. The cementum length decreased from 16 via 15 and 14, to 13 mm towards the apex. The cementum occupies only a part of the tooth that corresponds to $3 \cdot 10^{-4}$ of the tooth area shown in orange, see Fig. 4.

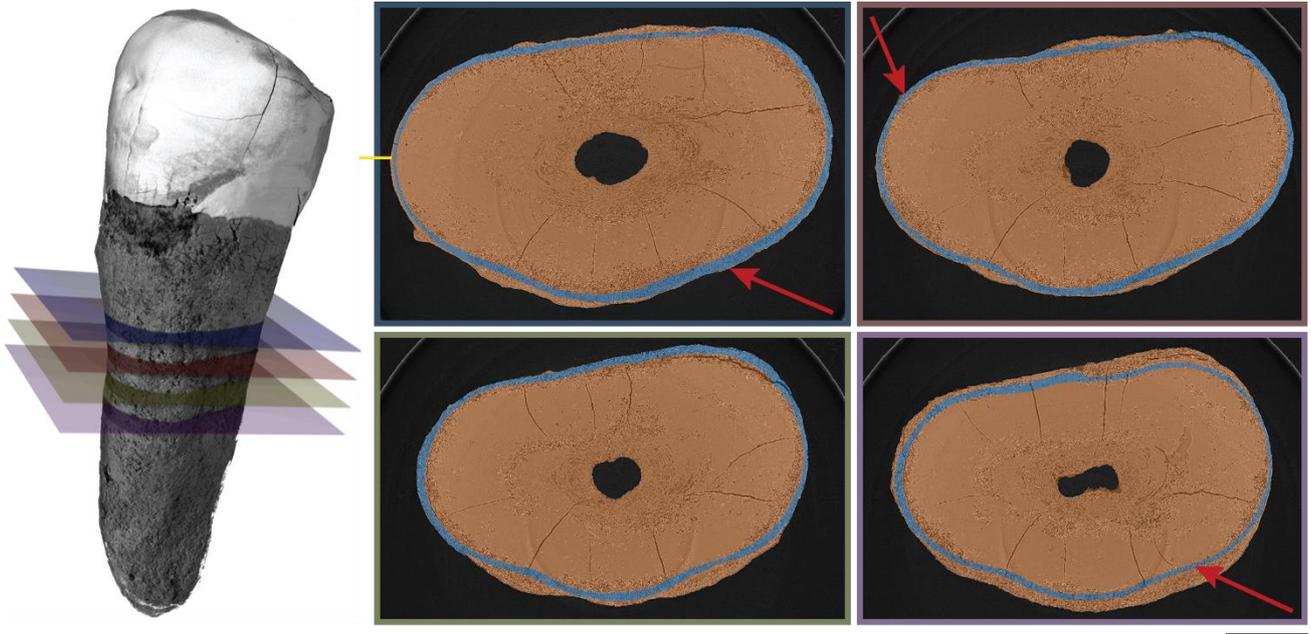

Figure 4. Hard X-ray data from an archeological human tooth of a 36-year-old woman. The planes on the left depict the four locations in the tooth root analyzed for incremental layer appearances. Incremental layers lie in the tooth cementum, which was manually segmented and is shown in blue color on the cross-sectional images indicated. The cross-section images were formed from integrating in the slice direction over a thickness of 55 µm. The length bar corresponds to 1 mm. A yellow marker shows the starting and end points of the representation in Fig. 5. The red-colored arrows point to the locations of annotated layers given in Fig. 6

### 3.3 Straighten the cementum of an archeological human tooth for improved imaging of annual layers

In previous work, we have shown that the visibility of incremental layers can automatically be improved by integrating along the layer direction [22]. Since incremental layers are curved and hardly aligned with the imaging direction, we automatically straightened the cementum before optimizing the integration direction. In addition, the straightening of the cementum reduces the images to the relevant field of view, which enables efficient visual inspection.

The straightening was based on extracting the centerline of the cementum segmentation, fitting a spline to 750-pixels long sections of the centerline, and resampling the image orthogonal to this spline using the code from R. Harkes [23]. Sections from the centerline were sampled in clockwise direction starting at the 9'o clock position, see yellow mark in the upper left cross-section of Fig. 4. The extracted straightened regions were oriented such that the tooth boundary appears at the image bottom.

After straightening the cementum, the incremental layers are mainly horizontally aligned. What remains to be done, is to align the layers across slices. For this purpose, we employed a method we recently proposed [22], where the integration direction is optimized to provide the strongest intensity variation orthogonal to the incremental layers, i.e. along vertically intensity profiles. The enhanced subregions, i.e., superimposed along the optimized direction, were then concatenated to allow inspecting the incremental layer progression.

Figure 5. shows the straightened and enhanced cementum for the cross-section closest to the enamel. Using this representation, the cementum can be visually examined for incremental layers and their progression. The straightened cementum of the four sections are shown in Videos 3 to 6. It is clearly observed that the appearance of incremental layers varies in presence, distances, and contrast along the cementum.

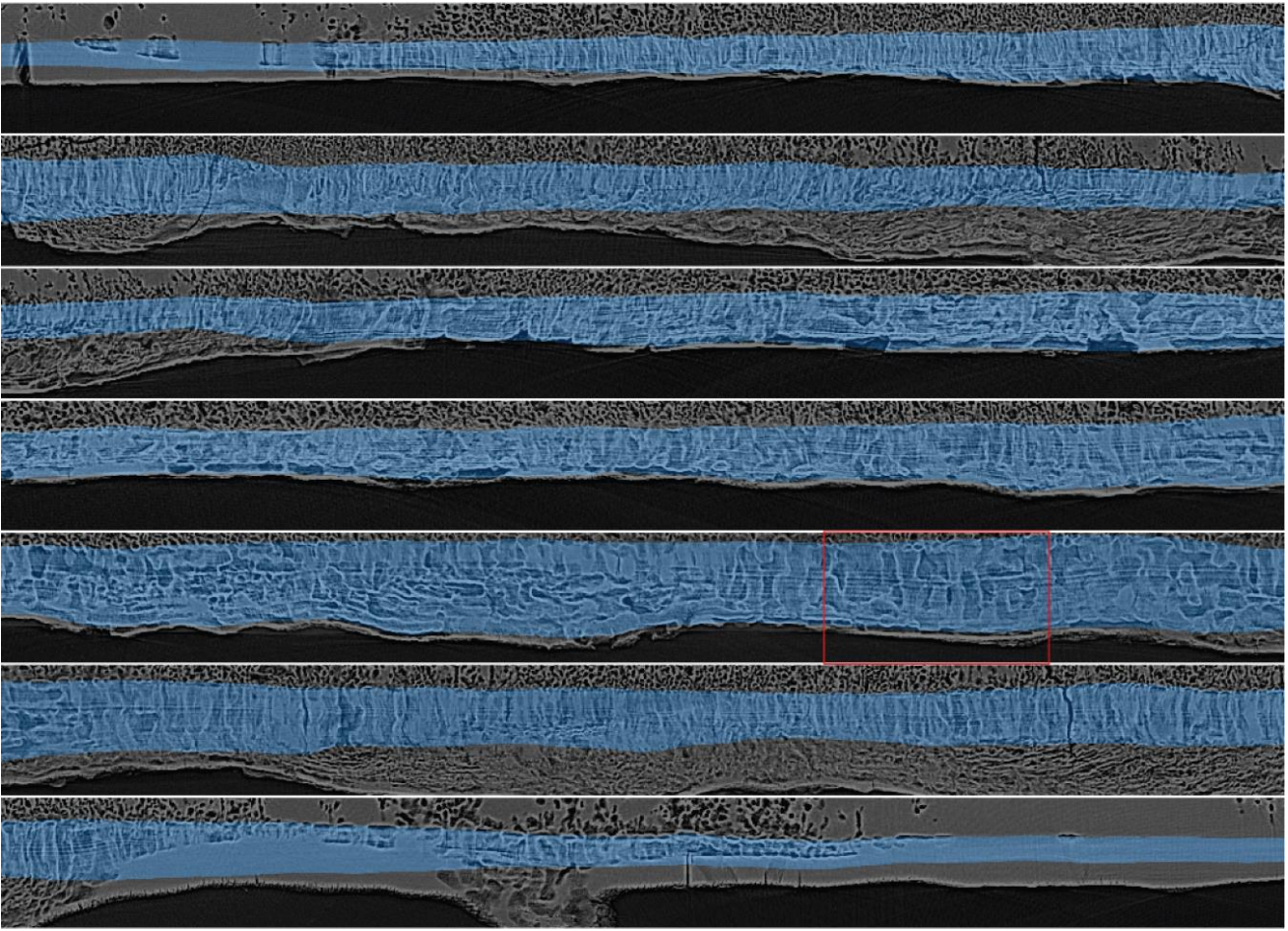

Figure 5. Straightened and enhanced cementum from plane marked in blue in Fig. 4, see cross section top left, showing (top to bottom) the result from clock-wise sampling the cross section starting at the 9 o'clock position, yellow marker in Fig. 4. This provides an overall visual impression of the cementum and incremental layers. The red-colored rectangle marks the location of region shown in the middle part of Fig. 6. The length bar corresponds to 100 µm.

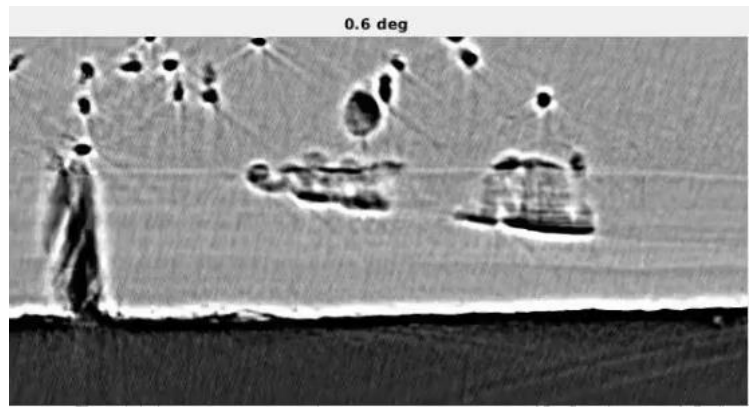

Video 3. Angular display of straightened and enhanced tooth cementum with a varying number of annual layers visible. The viewing angle is given; zero corresponds to the yellow mark in Fig. 4. The data correspond to the height step displayed in the top left of Fig. 4:  http://dx.doi.org/doi.number.goes.here

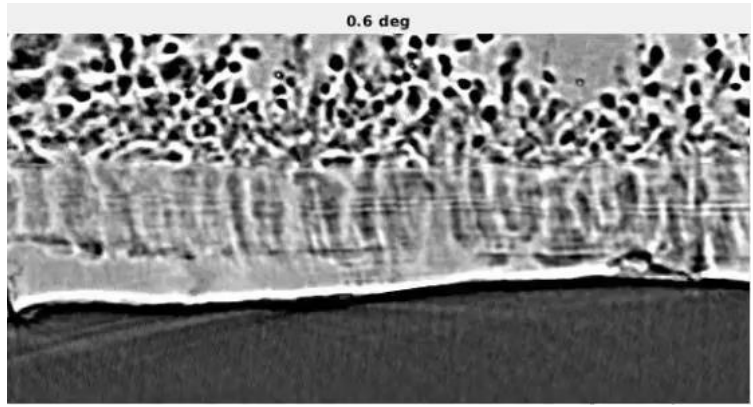

Video 4. Angular display of straightened and enhanced tooth cementum with a varying number of annual layers visible. The data correspond to the height step displayed in the top right of Fig. 4:  http://dx.doi.org/doi.number.goes.here

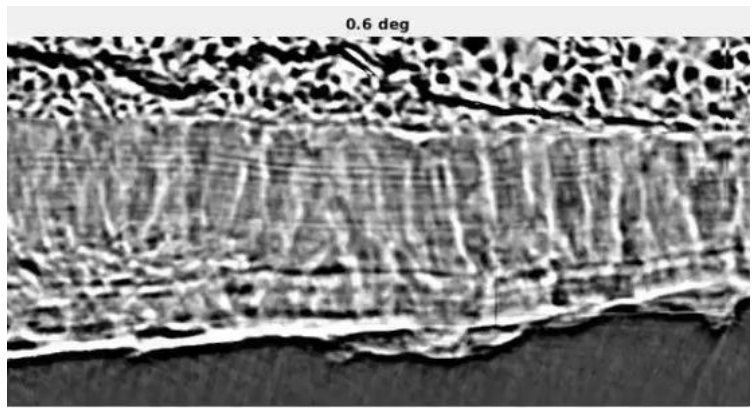

Video 5. Angular display of straightened and enhanced tooth cementum with a varying number of annual layers visible. The data correspond to the height step displayed at the bottom left of Fig. 4:  http://dx.doi.org/doi.number.goes.here

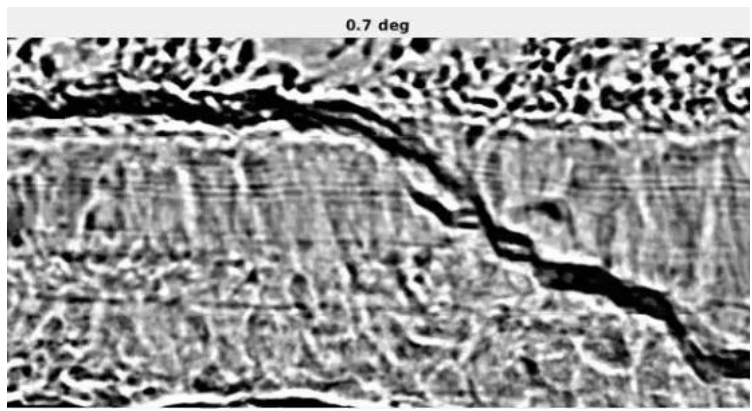

Video 6. Angular display of tooth cementum with a varying number of annual layers visible. The data correspond to the height step displayed at the cross-section lower right of Fig. 4:  http://dx.doi.org/doi.number.goes.here

### 3.4 Annotation of annual layers in cementum of a single human tooth

The bright incremental layers in selected, non-overlapping, enhanced regions were manually annotated to quantify the inter- and intra-variability of the layered structures. Four regions per height step were selected: (i) the regions following the yellow bar (9'clock position), (ii) the region with thickest cementum, and (iii) a selection by the annotator (C.T.).

Figure 6 shows the annotations for the region with the maximum number of incremental layers found (top image), the region with the thickest incremental layers (middle image), and the region with the thinnest incremental layers (bottom image). The incremental layers exhibited an average periodicity of 5.4 µm, see Table 1.

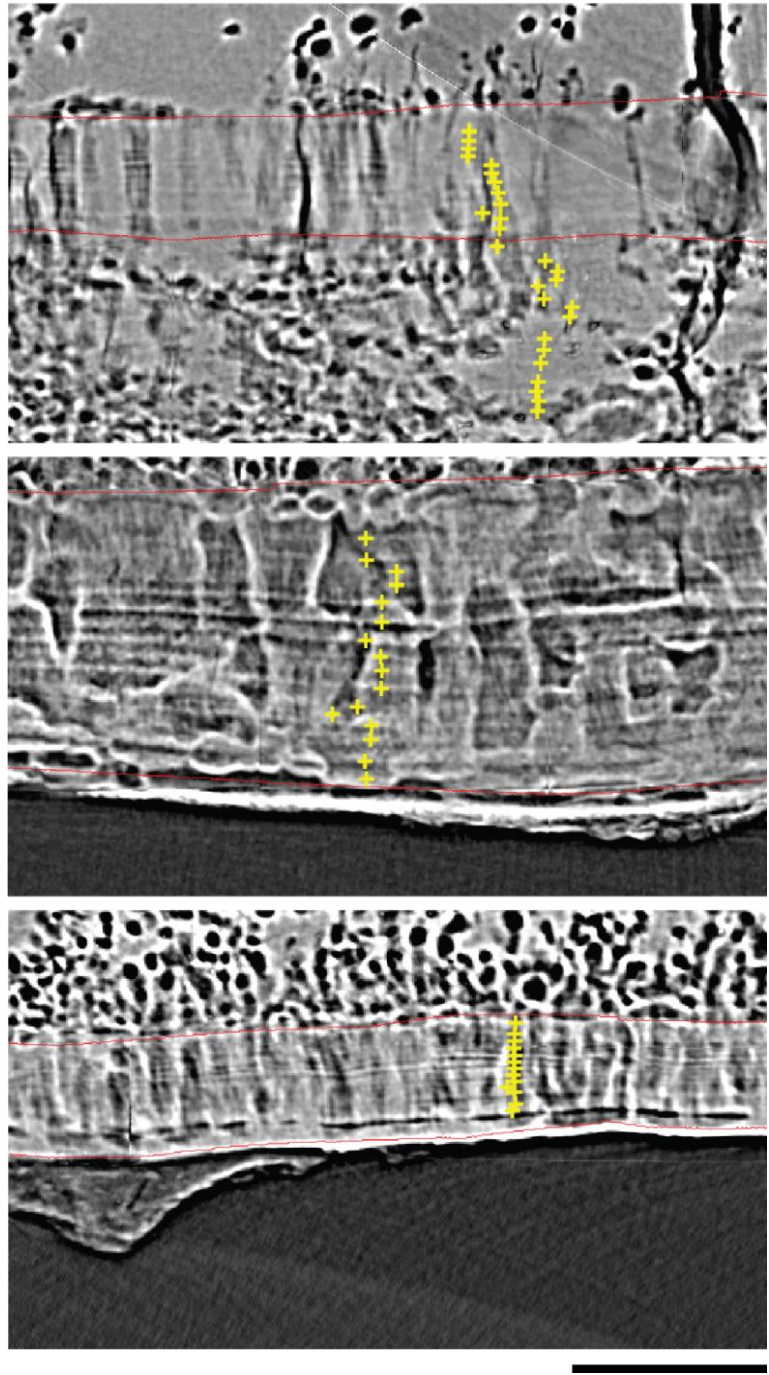

Figure 6. Cementum regions, optimized for integration direction, with location of manually detected bright incremental layers depicted by yellow-colored crosses. The borders of the cementum segmentation are shown by red contours. The red arrow head in the cross-section given in the lower right of Fig. 4 shows the position of the image on the top. It contains 27 incremental layers. These layers are also weakly visible between the pores in the outer region, a part excluded from the cementum segmentation. The middle image stems from the height step closest to the enamel, see red arrow head for the top left cross-section in Fig. 4. With 8.3 µm between incremental layers, this part shows the largest mean distance. The image

on the bottom belongs to the height step represented in the cross-section of Fig. 4 top right. Here, we found with 4.2 µm the smallest mean distance between incremental lines. The length bar corresponds to 100 µm.

Table 1. Mean distances, given in µm, between the detected incremental layers from the four height steps, see Fig. 4 (columns) and the three selected regions (rows). The error bars are represented via the standard deviations.

| Selection method | Height step closest to enamel, blue-colored | Height step orange-colored | Height step green-colored | Height step purple-colored | Mean value |
| --- | --- | --- | --- | --- | --- |
| "yellow bar" | 5.2 ± 2.4 | 4.7 ± 1.4 | 5.1 ± 1.8 | 5.4 ± 1.6 | 5.2 ± 1.8 |
| Thickest cementum | 6.1 ± 1.7 | 5.5 ± 1.3 | 5.0 ± 1.8 | 5.7 ± 1.9 | 5.6 ± 1.7 |
| Selected by C.T. | 5.7 ± 2.4 | 5.0 ± 1.8 | 5.4 ± 2.0 | 5.1 ± 1.9 | 5.3 ± 2.0 |
| Mean value | 5.7 ± 2.1 | 5.1 ± 1.5 | 5.2 ± 1.9 | 5.4 ± 1.8 | 5.4 ± 1.9 |

Table 2 lists the maximum number of manually detected incremental layers per height step and selection method. Although, one expects a constant number of annual layers throughout the tooth cementum, most of the numbers are substantially lower than the maximum value found, i.e. 27 layers.

Table 2. Maximum number of detected incremental layers.

| Selection method | Height step closest to enamel, blue-colored | Height step orange-colored | Height step green-colored | Height step purple-colored | Mean value |
| --- | --- | --- | --- | --- | --- |
| "yellow bar" | 15 | 13 | 22 | 18 | 17.0 |
| Thickest cementum | 25 | 21 | 18 | 20 | 21.0 |
| Selected by C.T. | 16 | 17 | 21 | 27 | 20.0 |
| Mean value | 19.5 | 18.0 | 19.8 | 21.8 | 19.8 |

### 3.5 Correlating tomography data with optical micrographs

Traditionally, optical microscopy is used to examine incremental layers on tooth sections of up to 100 µm thickness [3]. In order to identify the advantages of each technique and to identify any possible advantage by their combination, we imaged human tooth slides using optical and hard X-ray tomography. Because the optical data contain only two dimensions, a direct registration and comparison is challenging. The projective nature of the optical micrographs yields better contrast than found in the individual virtual slices of hard X-ray tomography. As a consequence, we improved the contrast of the images in both modalities. Optical data were enhanced by adaptive histogram equalization of the intensities. The representation of the hard X-ray tomograms of the cementum region was improved integrating along the incremental layer direction as determined by the optimization procedure described in section 3.3 [22].

In a first step, the overview optical image was registered with the hard X-ray tomogram to globally align the images from the two modalities. Note, the overlapping local optical scans were stitched, such that they were consistent with the overview optical scan. Hence, the local regions of the hard X-ray tomogram corresponding to the local optical scan could be extracted. In a final step, the alignment was fine-tuned by rigidly registering the enhanced local images based on nine landmark pairs manually selected by C.T., which were mainly located on the eruption layer. A characteristic example is shown in Fig. 7. Here, one can clearly recognize corresponding micro-anatomical features in the tooth cementum and beyond. Visual inspection of the common area allows for the detection of the incremental layers. Hence, the layers visible in hard X-ray tomography relate to incremental layers detected via optical microscopy.

The observation of the annual layer structures in the center of this corresponding region, however, led to subtle distinctions: for the X-ray data, we detected 18 layers with a mean distance of 6.0 µm, whereas the optical image contained 30 layers with a mean distance of 2.4 µm. During annotation of the optical image, we observed that dark layers were generally thinner than the bright ones. These sub-micrometer-thin layers were obviously not resolved by the hard X-ray tomography approach.

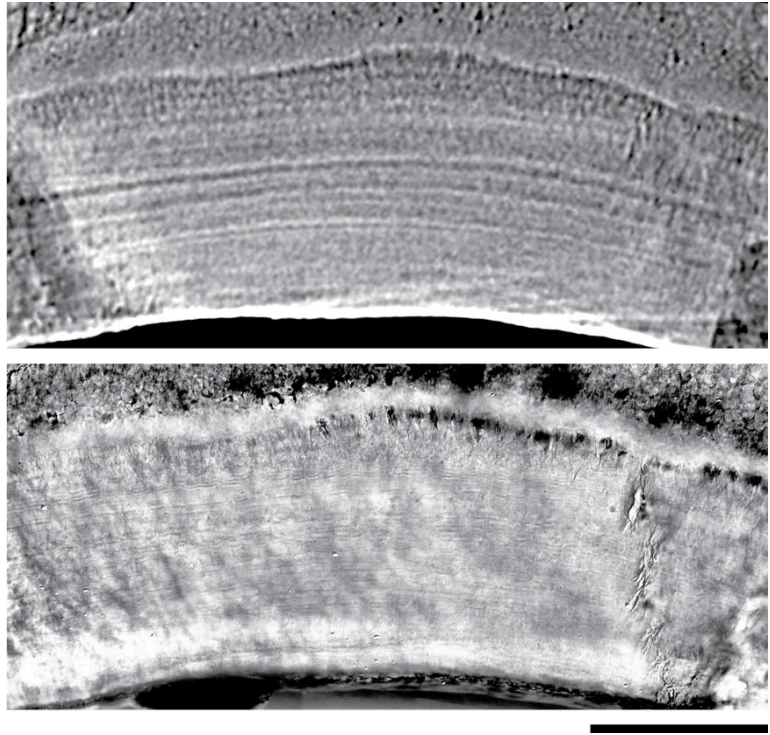

Figure 7. Comparison of incremental layer appearance for same region using hard X-ray tomography (top image) and optical imaging (bottom image). The length bar corresponds to 100 μm.

### 3.6 Layered structures in buffalo tooth

In contrast to human maxillary premolars with a single root, the buffalo tooth investigated has four roots of unequal shape. Its size was much larger than the human tooth studied. Therefore, the straightening and enhancement, described in section 3.3, was only be applied to promising, user-selected sub-volumes. For this purpose, we visually inspected twelve cross-sections, super-positioned 20 tooth cementum slices, and applied the enhancement. Furthermore, we inspected the intensity variations across the slices to identify cementum regions with highest benefit from the optimization of the integration direction. This protocol gave rise to incremental layers in the distal region of the buffalo tooth just below the enamel, see red-colored arrow in Fig. 8. This enhancement was achieved by integrating the 20 slices after a 23.8 degrees rotation around the horizontal axis. The integration over 10, 30 or even 50 slices yielded visually less sharp features. Starting from the selected location, the incremental layers were visible over a range of at least 1 mm towards the apex.

Nine bright incremental layers were detected for this region, as indicated in the lower right image of Fig. 8. The mean distance between the layers was found to be 45.9 μm. This result does not only show that hard X-ray tomography can be helpful to detect incremental layers in bovids, but also that their thickness is an order of magnitude larger than in human tooth cementum.

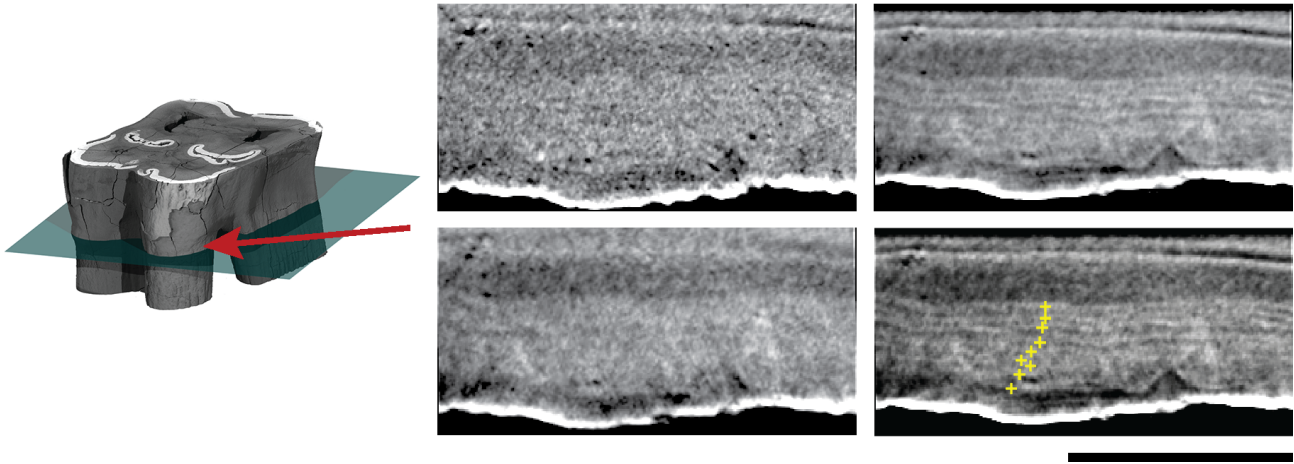

Figure 8. On African buffalo, tooth cementum accumulates on the region below the enamel, see red-colored arrowhead. This arrow points to the region, where the tomographic slices have been acquired. Individual slices about 6 µm thick show weak contrast, top left. The integration of 20 slices is hardly helpful, see lower left image. The integration along the optimized direction, however, brings the layered structure to light, see right images. In the lower image, the nine layers are annotated by the yellow-colored crosses. The length of the bar corresponds to 1 mm.

## 4. DISCUSSION

### 4.1 Analysis of tree's annual growth rings

Although hard X-ray tomographic imaging with micrometer resolution is well suited to study the annual layers in wood, its application to wood is commonly questionable. There are a few exceptions, which might be considered as niche research activities. Unique wooden objects, generally opaque, could be three-dimensionally characterized, see for example [24]. Water uptake or chemical modifications are applications, where microtomography supports the quantitative characterization of wood, see for example [25, 26]. It can also be used to quantify the quality of wood pellets including pore and multifractal structure [27]. Basic research activities, however, such as hard X-ray micro-densitometry on annually resolved tree-ring samples have gained interest in last-millennium paleoclimatology through density parameters including the maximum latewood density, see review article of Björklund *et al.* [28].

### 4.2 Three-dimensional imaging of annual layers in human tooth cementum

Imaging, enhancing, and inspecting the cementum of entire cross-sections over several height steps of an individual tooth made it possible to find a region, where the number of detected incremental layers was similarly high as the expected one. Studies sampling only few cementum regions are likely to miss such valuable observations, which then would lead to an underestimation of age-at-death. An efficient path might be searching for the thickest cementum layers and count the layers only in this volume, see Table 2.

The spacing between incremental layers varied substantially and was on average larger than previously reported, see for example [3]. For 32 archeological human teeth, G. Mani-Caplazi reported a mean value of 2.9 µm, which is almost half of the value found for the tomographic imaging in the present study. This substantial discrepancy between optical micrographs and tomographic data might be attributed to the image modality. The sub-micrometer-thin features detected in the optical micrographs could not be resolved with tomographic imaging using sub-micrometer voxel lengths. In order to proof this hypothesis, one should image tooth slices with nano-tomography, as for example available at the European Synchrotron Radiation Facility in France [29], and using polarization microscopy.

The question is whether the maximum number of layers identified corresponds to the age-at-death. Maria Magdalena Scherb reached an age between 36 and 37 years, and we identified 27 annual layers. As the mean tooth eruption age of the first premolar in females corresponds to 11.63 years [30], the fit is almost perfect. It should be noted that a small body-mass-index gives rise to earlier tooth eruption [30]. Therefore, we can reasonably expect that the eruption of the first premolar of Maria Magdalena Scherb occurred at an age between nine and ten years. At the other locations investigated, however, we identified less than 27 incremental layers and the related age-at-death would be underestimated. Therefore, the numbers found are rather a lower limit for the estimation of the age-at-death.

### 4.3 Imaging of layered structures in bovid's teeth

The successful imaging of the layered structures in the tooth cementum of the African buffalo could be regarded as a milestone towards the identification of previous hunting seasons. As the periodicity of the annual layers is close to 50 µm, the spatial resolution of currently available laboratory-based microtomography systems should be satisfactory to determine the month-at-death [31]. Besides the spatial resolution, the density resolution is critical [32]. Therefore, the structural and chemical differences between the dark and bright layers should be understood in detail. So far, we can only speculate on the origin of the layered microstructures in bovid teeth. The formation, however, might have similarities to the layers in human tooth cementum, because the number of detected incremental layers coincides with the age-at-death of the animal, see [18]. The preliminary results of the current study suggest that the detected layers can be described as annual rings with a thickness one order of magnitude thicker than in human teeth. A proof of this suggestion requires a series of experiments, where hard X-ray microtomography should play a key role.

## 5. CONCLUSIONS

Tomographic imaging using hard X rays is a powerful method for investigating microscopic anatomical features in a macroscopic object avoiding physical slicing. Annual layers, known from wood with a periodicity of around a millimeter, can also be identified in animal and human tissues, namely in tooth cementum. Tooth cementum contains a wealth of information on age, stress periods such as pregnancy, health issues, and (restricted) nutrition. Because unique objects should remain as an integrated whole, the sample preparation only relies on holder selection. Image acquisition can include data of terabyte size and are therefore time-consuming. Data reconstruction and analyses have to be automatic and based on sophisticated tools known from computational sciences. Interdisciplinary teams are a prerequisite for successful data collection and quantification.

## ACKNOWLEDGEMENTS

The authors are especially thankful to Meinhard F. Müller, Freienstein, Switzerland for the valuable gift, i.e. the wooden disks, and the related preparation for the imaging. The authors are grateful to Synchrotron SOLEIL for the provision of beamtime at the ANATOMIX beamline within the frame of proposals 20200712 and 20210554. ANATOMIX is an Equipment of Excellence (EQUIPEX) funded by the Investments for the Future program of the French National Research Agency (ANR), project NanoimagesX, grant no. ANR-11-EQPX-0031. The authors thank the company Carl Zeiss Microscopy GmbH, Oberkochen, Germany for the possibility to image the buffalo tooth three-dimensionally, as represented in Fig. 8 on the left.

## REFERENCES

[1] Jean, B., and Pierre, D., "Applications of computed tomography (CT) scanning technology in forest research: a timely update and review," Canadian Journal of Forest Research **49**(10), 1173-1188 (2019).

[2] Naji, S., Rendu, W., and Gourichon, L., [Dental Cementum in Anthropology] Cambridge University Press, Cambridge (2022).

[3] Mani-Caplazi, G., Schulz, G., Deyhle, H., Hotz, G., Vach, W., Wittwer-Backofen, U., and Müller, B., "Imaging of the human tooth cementum ultrastructure of archeological teeth, using hard X-ray microtomography to determine age-at-death and stress periods," Proceedings of SPIE **10391**, 103911C (2017).

[4] Hotz, G., and Steinke, H., "Knochen, Skelette, Krankengeschichten. Spitalfriedhof und Spitalarchiv - zwei sich ergänzende Quellen," Basler Zeitschrift für Geschichte und Altertumskunde **112**, 105-138 (2012).

[5] Karakostis, F. A., Hotz, G., Scherf, H., Wahl, J., and Harvati, K., "Occupational manual activity is reflected on the patterns among hand entheses," Am J Phys Anthropol **164**(1), 30-40 (2017).

[6] Karakostis, F. A., Hotz, G., Scherf, H., Wahl, J., and Harvati, K., "A repeatable geometric morphometric approach to the analysis of hand entheseal three-dimensional form," Am J Phys Anthropol **166**(1), 246-260 (2018).

[7] Weitkamp, T., Scheel, M., Giorgetta, J. L., Joyet, V., Le Roux, V., Cauchon, G., Moreno, T., Polack, F., Thompson, A., and Samama, J.-P., "The tomography beamline ANATOMIX at Synchrotron SOLEIL," Journal of Physics: Conference Series **849**, 012037 (2017).

[8] Desjardins, K., Scheel, M., Giorgetta, J.-L., Weitkamp, T., Menneglier, C., and Carcy, A., "Design of indirect X-ray detectors for tomography on the ANATOMIX beamline," Proceedings of Tenth Mechanical Engineering Design of Synchrotron Radiation Equipment and Instrumentation, MEDSI2018, 355-357 (2018).

[9] Du, M., Vescovi, R., Fezzaa, K., Jacobsen, C., and Gürsoy, D., "X-ray tomography of extended objects: a comparison of data acquisition approaches," J. Opt. Soc. Am. A **35**(11), 1871-1879 (2018).

[10] Miettinen, A., Oikonomidis, I. V., Bonnin, A., and Stampanoni, M., "NRStitcher: non-rigid stitching of terapixel-scale volumetric images," Bioinformatics **35**(24), 5290-5297 (2019).

[11] Vescovi, R., Du, M., Andrade, V. d., Scullin, W., Gürsoy, D., and Jacobsen, C., "Tomosaic: efficient acquisition and reconstruction of teravoxel tomography data using limited-size synchrotron X-ray beams," Journal of Synchrotron Radiation **25**(5), 1478-1489 (2018).

[12] Vescovi, R. F. C., Cardoso, M. B., and Miqueles, E. X., "Radiography registration for mosaic tomography," Journal of Synchrotron Radiation **24**(3), 686-694 (2017).

[13] Rodgers, G., Schulz, G., Deyhle, H., Kuo, W., Rau, C., Weitkamp, T., and Müller, B., "Optimizing contrast and spatial resolution in hard x-ray tomography of medically relevant tissues," Applied Physics Letters **116**(2), 023702 (2020).

[14] Boin, M., and Haibel, A., "Compensation of ring artefacts in synchrotron tomographic images," Opt. Express **14**(25), 12071-12075 (2006).

[15] Mirone, A., Brun, E., Gouillart, E., Tafforeau, P., and Kieffer, J., "The PyHST2 hybrid distributed code for high speed tomographic reconstruction with iterative reconstruction and a priori knowledge capabilities," Nuclear Instruments and Methods in Physics Research Section B: Beam Interactions with Materials and Atoms **324**, 41-48 (2014).

[16] Marone, F., and Stampanoni, M., "Regridding reconstruction algorithm for real-time tomographic imaging," Journal of Synchrotron Radiation **19**(6), 1029-1037 (2012).

[17] Gürsoy, D., De Carlo, F., Xiao, X., and Jacobsen, C., "TomoPy: a framework for the analysis of synchrotron tomographic data," Journal of Synchrotron Radiation **21**(5), 1188-1193 (2014).

[18] Bunn, H. T., Bartram, L. E., and Kroll, E. M., "Variability in bone assemblage formation from Hadza hunting, scavenging, and carcass processing," Journal of Anthropological Archaeology **7**(4), 412-457 (1988).

[19] Bunn, H. T., "Large ungulate mortality profiles and ambush hunting by Acheulean-age hominins at Elandsfontein, Western Cape Province, South Africa," Journal of Archaeological Science **107**, 40-49 (2019).

[20] Müller, B., Deyhle, H., Lang, S., Schulz, G., Bormann, T., Fierz, F., and Hieber, S., "Three-dimensional registration of tomography data for quantification in biomaterials science," International Journal of Materials Research **103**(2), 242-249 (2012).

[21] Paganin, D., Mayo, S. C., Gureyev, T. E., Miller, P. R., and Wilkins, S. W., "Simultaneous phase and amplitude extraction from a single defocused image of a homogeneous object," Journal of Microscopy **206**(1), 33-40 (2002).

[22] Tanner, C., Rodgers, G., Schulz, G., Osterwalder, M., Mani-Caplazi, G., Hotz, G., Scheel, M., Weitkamp, T., and Müller, B., "Extended-field synchrotron microtomography for non-destructive analysis of incremental lines in archeological human teeth cementum," Proceedings of SPIE **11840**, 1184019 (2021).

[23] Harkes, R., [Straighten an image in Matlab], (2021).

[24] Mizuno, S., Torizu, R., and Sugiyama, J., "Wood identification of a wooden mask using synchrotron X-ray microtomography," Journal of Archaeological Science **37**(11), 2842-2845 (2010).

[25] Klingner, R., Sell, J., Zimmermann, T., Herzog, A., Vogt, U., Graule, T., Thurner, P., Beckmann, F., and Müller, B., "Wood-derived porous ceramics via infiltration of SiO2-sol and carbothermal reduction," Holzforschung **57**(4), 440-446 (2003).

[26] Moghaddam, M. S., Bulcke, J. V. d., Wålinder, M. E. P., Claesson, P. M., Acker, J. V., and Swerin, A., "Microstructure of chemically modified wood using X-ray computed tomography in relation to wetting properties," Holzforschung **71**(2), 119-128 (2017).

[27] Srocke, F., Han, L., Dutilleul, P., Xiao, X., Smith, D. L., and Mašek, O., "Synchrotron X-ray microtomography and multifractal analysis for the characterization of pore structure and distribution in softwood pellet biochar," Biochar **3**(4), 671-686 (2021).

[28] Björklund, J., von Arx, G., Nievergelt, D., Wilson, R., Van den Bulcke, J., Günther, B., Loader, N. J., Rydval, M., Fonti, P., Scharnweber, T., Andreu-Hayles, L., Büntgen, U., D'Arrigo, R., Davi, N., De Mil, T., Esper, J., Gärtner, H., Geary, J., Gunnarson, B. E., Hartl, C., Hevia, A., Song, H., Janecka, K., Kaczka, R. J., Kirdyanov, A. V., Kochbeck, M., Liu, Y., Meko, M., Mundo, I., Nicolussi, K., Oelkers, R., Pichler, T., Sánchez-Salguero, R., Schneider, L., Schweingruber, F., Timonen, M., Trouet, V., Van Acker, J., Verstege, A., Villalba, R., Wilmking, M., and Frank, D., "Scientific merits and analytical challenges of tree-ring densitometry," Reviews of Geophysics **57**(4), 1224-1264 (2019).

[29] Khimchenko, A., Bikis, C., Pacureanu, A., Hieber, S. E., Thalmann, P., Deyhle, H., Schweighauser, G., Hench, J., Frank, S., Müller-Gerbl, M., Schulz, G., Cloetens, P., and Müller, B., "Hard X-ray nano-holotomography: Large-scale, label-free, three-dimensional neuroimaging beyond optical limit," Advanced Science **5**(6), 1700694 (2018).

[30] Khan, A. S., Nagar, P., Singh, P., and Bharti, M., "Changes in the sequence of eruption of permanent teeth; Correlation between chronological and dental age and effects of body mass index of 5-15-year-old schoolchildren," Int J Clin Pediatr Dent **13**(4), 368-380 (2020).

[31] Migga, A., Schulz, G., Rodgers, G., Osterwalder, M., Tanner, C., Blank, H., Jerjen, I., Salmon, P., Twengström, W., Scheel, M., Weitkamp, T., Schlepütz, C. M., Bolten, J. S., Huwyler, J., Hotz, G., Madduri, S., and Müller, B., "Comparative hard x-ray tomography for virtual histology of zebrafish larva, human tooth cementum, and porcine nerve," Journal of Medical Imaging **9**(3), 031507 (2022).

[32] Thurner, P., Beckmann, F., and Müller, B., "An optimization procedure for spatial and density resolution in hard X-ray micro-computed tomography," Nuclear Instruments and Methods in Physics Research Section B: Beam Interactions with Materials and Atoms **225**(4), 599-603 (2004).